\pdfoutput=1 
\documentclass[preprint, 3p]{elsarticle}
\usepackage{lineno,hyperref}
\usepackage{array}
\usepackage{graphicx}
\usepackage{adjustbox}
\usepackage{subcaption}
\usepackage{float}
\usepackage{amsmath}
\usepackage{xcolor}
\usepackage{graphicx}
\usepackage{dcolumn}

\usepackage[utf8]{inputenc}
\usepackage[T1]{fontenc}
\usepackage{mathptmx}
\journal{Elsevier}

\begin{document}
\begin{frontmatter}

\title{Correlating Local Chemical and Structural Order Using Geographic Information Systems-Based Spatial Statistics} 

\author[ad1]{Michael Xu\fnref{1}}
\address[ad1]{Massachusetts Institute of Technology, Cambridge, MA 02139, USA}
\fntext[1]{These authors contributed equally.}

\author[ad1]{Abinash Kumar\fnref{1}}

\author[ad1]{James M. LeBeau\texorpdfstring{\corref{mycorrespondingauthor}}}

\cortext[mycorrespondingauthor]{Corresponding author, E-mail: {\tt lebeau@mit.edu}}

\begin{abstract}

Analysis of nanoscale short-range chemical and/or structural order via (scanning) transmission electron microscopy (S/TEM) imaging is fundamentally limited by projection of the three dimensional sample, which averages informational along the beam direction. Extracting statistically significant spatial correlations between the structure and chemistry determined from these two-dimensional datasets thus remains challenging. Here, we apply methods commonly used in Geographic Information Systems (GIS) to determine the spatial correlation between measures of local chemistry and structure from atomic-resolution STEM imaging of a compositionally complex relaxor, Pb(Mg$_{1/3}$Nb$_{2/3}$)O$_{3}$ (PMN). The approach is used  to determine the type of ordering present and to quantify the spatial variation of chemical order, oxygen octahedral distortions, and oxygen octahedral tilts.  The extent of autocorrelation and inter-feature correlation among these short-range ordered regions are then evaluated through a spatial covariance analysis, showing correlation as a function of distance. The results demonstrate that integrating GIS tools for analyzing microscopy datasets can serve to unravel subtle relationships among chemical and structural features in complex materials that can be hidden when ignoring their spatial distributions.

\end{abstract}
\begin{keyword}
scanning transmission electron microscopy
\sep geographic information systems
\sep short-range order
\end{keyword}

\end{frontmatter}

%
%

\section{Introduction}





In many materials, short-range chemical and structural order are key to controlling material properties. For example, careful control of composition and/or processing conditions can yield varying degrees of short-range order (SRO), which can dramatically impact properties ranging from mechanical behavior \citep{Miracle2015CriticalMaterials, Miracle2017AConcepts, Diao2017FundamentalOverview} to dielectric response \citep{Randall_1990, Takesue_Fujii_Ichihara_Chen_Tatemori_Hatano_1999, ning_ross_antisite2021}. Analysis of SRO in such materials typically relies on X-ray and/or neutron diffraction, using Bragg peaks to determine the global structure and diffuse scattering to derive information about the nature of chemical and structural ordering \citep{Proffen_2000, Proffen_Petkov_Billinge_Vogt_2002}. 


To study SRO directly at the atomic scale, scanning transmission electron microscopy (STEM) offers direct, simultaneous imaging of both local chemistry and structure. Specifically, unit cell level measurements of order/disorder can be derived from images sensitive to the type and number of atoms present (by annular dark-field STEM and/or spectroscopy) \citep{Pennycook_Boatner_1988, Browning_Chisholm_Pennycook_1993, DAlfonso_Freitag_Klenov_Allen_2010} or the location of light atoms (by annular bright-field \citep{Findlay2010} or differential phase contrast STEM \citep{lazic2016}). Thus, the distribution of order and distortion can be quantified, enabling correlations to be measured and related to material performance \citep{Danaie_asitestem_2016, cabralgradient2018, Ding2019, chenma2021}. Quantifying relationships between SRO features, however, can benefit from the application of spatial statistics to understand spatial correlations and their significance.



Correlating chemical and structural SRO captured with the electron microscope is conceptually similar to geographic analysis in that it involves understanding spatial relationships between complex phenomena. In microscopy, features such as chemistry or distortion can be measured across atom columns or entire grains, and in geographic analysis population or income can be measured across individual towns to whole countries. While the process of drawing conclusions from data in both these cases can be similar, the use of formalized spatial statistics is nearly ubiquitous in geographic analysis. These methods, collectively incorporated into database and analysis tools known as Geographic Information Systems (GIS), provide ways to aggregate a wide variety of data from surveys, sampling, and remote sensing in order to analyze spatial patterns, identify clustering and dispersion, and map spatio-temporal relationships \citep{Burrough_McDonnell_1998, Longley_2005, spdep}. Since pertinent topics in materials science such as chemical and structural distribution, clustering, ordering, and transformation are invariably coupled in space and time, deciphering these phenomena can benefit from the rigorous tools available from applying spatial statistics. 


Analyzing short-range order in relaxor ferroelectrics serves as a prototypical example application space for spatial statistics. In these materials, order and disorder have been shown to induce a wide variety of changes in material behavior, such as piezoelectric and dielectric response \citep{EricCross1987,Ahn2016}, and significant effort has been made to connect global and local chemistry and structure to their performance. For example, the authors have previously explored the Pb(Mg$_{1/3}$Nb$_{2/3}$)O$_{3}$ (PMN) system, a globally cubic (space group $Pm\bar{3}m$) perovskite oxide with Pb on the A sub-lattice and a mixed B sub-lattice occupied by Mg and Nb. In addition, this material possesses short-range chemically ordered regions (CORs) arising from B-site ordering of Mg and Nb \citep{Krause_Cowley_Wheatley_1979, Randall_1990, bhalla_1990, Kopecky2016, cabralgradient2018, Eremenko2019, kumar_2021pmn}, ordered oxygen octahedral distortion regions (ODRs), and ordered oxygen octahedral tilt regions (OTRs) \citep{Rosenfeld_Egami_1995, Krogstad2018, kumar_2021pmn}. These structural and chemical heterogeneities exist at the local level and are thought to govern the relaxor behavior in these materials \citep{EricCross1987,Ahn2016, Randall_1990,bhalla_1990,Kopecky2016,Eremenko2019}. 





In this study, GIS-based spatial statistics techniques are applied to atomic resolution chemistry and structure ``maps'' to uncover local chemical and structural correlations in the relaxor PMN. Specially, the presence of CORs, ODRs, and OTRs are used to demonstrate  quantification of order and correlative analysis provided by spatial statistics. Using measures such as the Moran's I statistic, chemical and structural order are identified and quantified using a near-neighbor shell approach similar to the Warren-Cowley parameter \citep{cowley1950cuau, cowley1950theory}, yet formulated for a two-dimensional real space image. Based on the average ordering behavior, the Local Moran's I statistic is used to map statistically significant order and disorder by means of a null hypothesis test. Finally, spatial covariance is  measured using the codispersion coefficient to quantify the relationship between chemical and structural order, demonstrating the statistical significance of correlation between chemistry and structure in this material. Ultimately, these results demonstrate how incorporating these methods into electron microscopy data analysis can provide direct insights into chemistry-structure relationships in materials exhibiting SRO.


\section{Materials and Methods}
PMN and PMN-30PT single crystals were grown using the high-temperature flux method \citep{Zhang2012, CABRAL2018b} and were prepared along $\left<110\right>$ for electron microscopy using mechanical wedge polishing followed by Ar$^+$ ion milling at cryogenic temperatures using a Fischione 1051 ion mill.  Simultaneous annular dark-field (ADF) and integrated differential phase contrast (iDPC) STEM imaging was performed with a probe-corrected Thermo Fisher Scientific Themis Z S/TEM operated at 200 kV and a convergence semi-angle of 18 mrad. ADF and iDPC images were acquired with collection angles of 25-153 mrad and 6-24 mrad, respectively. The Revolving STEM (RevSTEM) method was applied to correct the image distortions arising from sample drift using 20 $2048 \times 2048$ pixel images collected with a 90$^\circ$ scan rotation between each successive frame \citep{revstem}.

A Python-based 2D Gaussian atom column fitting and analysis method was performed to extract atom column intensity, position, normalized intensity of the Mg/Nb sub-lattice (with respect to nearest-like neighbors), oxygen-oxygen distance, and oxygen-oxygen tilt angle with respect to the $\{002\}$-planes containing Mg, Nb, and O. These data were exported to R, which was then used to conduct GIS spatial statistics analyses, tests, and visualizations using the \textit{Spatial Dependence: Weighting Schemes, Statistics} (\textbf{spdep} \citep{Bivand2018}), \textit{SpatialPack: Tools for Assessment the Association Between Two Spatial Processes} (\textbf{SpatialPack} \citep{vallejos2020}), and \textit{tmap: thematic maps in R} (\textbf{tmap} \citep{tennekes_tmap}) packages. An example script applying the analysis can be downloaded at https://github.com/LeBeauGroup/GISMicroscopy.

\section{Results and Discussion}



\begin{figure}
    \centering
    \includegraphics[width=3in]{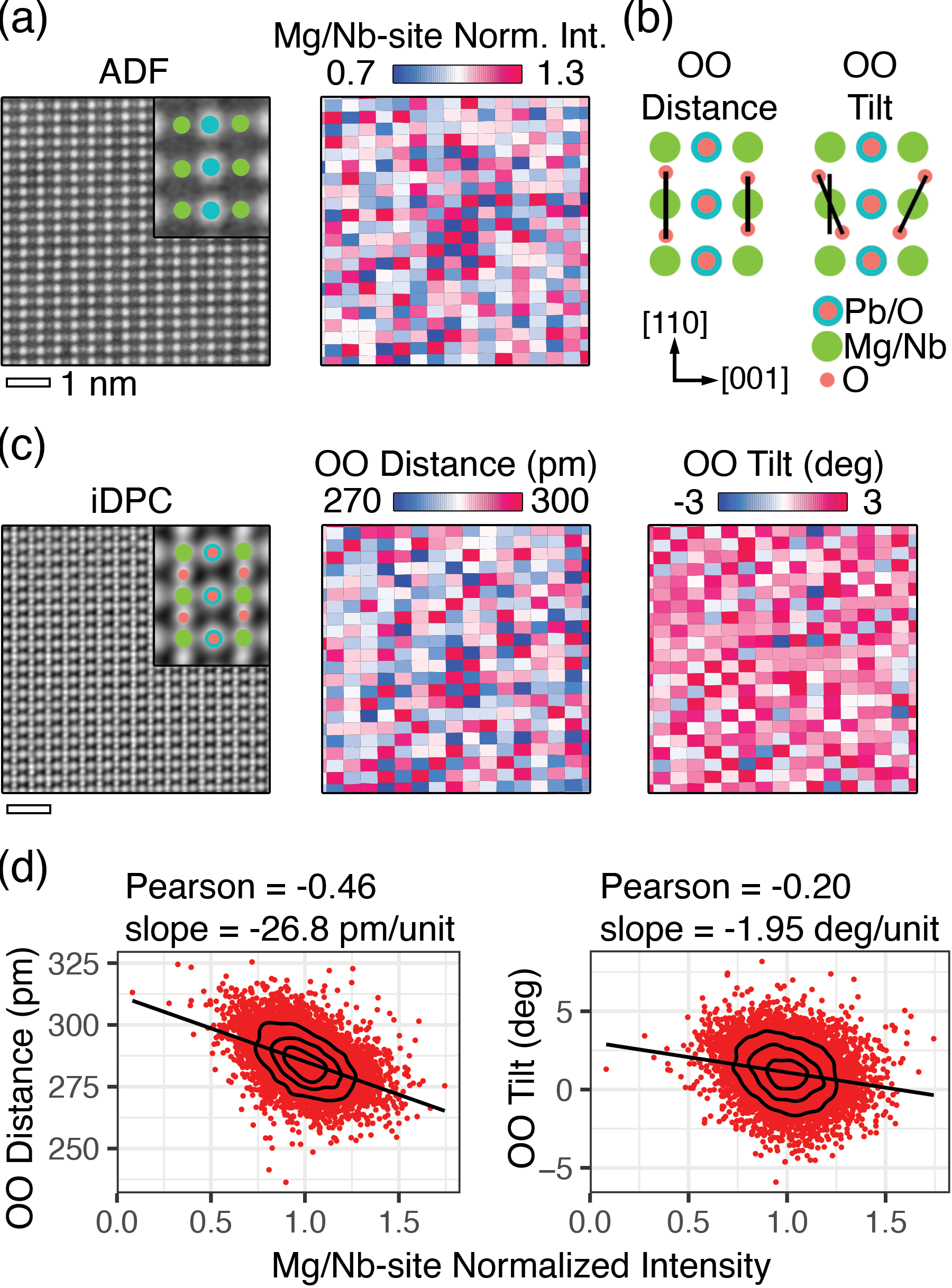}
    \caption{(a) Mg/Nb-site normalized intensity calculated from STEM ADF images with inset identifying Pb and Mg/Nb atom columns, (b) schematic of oxygen-oxygen distance and tilt measurement, and (c) corresponding distance and tilt maps extracted from STEM iDPC images with inset identifying Pb/O, Mg/Nb, and O atom columns. (d) Dependence of distance and tilt on Mg/Nb-site normalized intensity. Each box represents one unit cell.}
    \label{fig:intro}
\end{figure}

Simultaneously acquired ADF and iDPC images of PMN, Figure~\ref{fig:intro}a,c, allow the local chemical and structural information to be extracted using the intensity and position of atom columns \cite{kumar_2021pmn}. The Mg/Nb sub-lattice positions exhibit significant variation in the ADF image due to inhomogeneous distribution of these elements along the atom columns. As shown in Figure~\ref{fig:intro}a, darker and brighter atom columns are evident and highlighted by the Mg/Nb-site normalized intensity map \cite{cabralgradient2018}. From measurements using the iDPC image (Figure~\ref{fig:intro}c), the oxygen octahedral distortion and oxygen octahedral tilt (defined in Figure~\ref{fig:intro}b) locally alternate between large/small and positive/negative, respectively.


Combining the intensity measurements from ADF with oxygen octahedral distortion measurements from iDPC, the Mg/Nb-site normalized intensities are linearly correlated with oxygen-oxygen distortion at each unit cell with a Pearson correlation coefficient of 0.4, as shown in Figure~\ref{fig:intro}d . Likewise, the oxygen-oxygen tilts show weak correlation (Pearson coefficient of 0.2) with the Mg/Nb-site normalized intensity. In line with the relatively low Pearson correlation coefficients, the high degree of scatter indicates a wide distribution of correlated and uncorrelated unit cells, likely due to partial correlation and projection of the structure along the beam direction. Quantifying the degree of correlation beyond a simple linear fit necessitates using spatial statistics. 

\subsection{Determining the Type of Order}

Correlating local/unit-cell level variations in chemistry and structure begins with determination of the type of ordering, i.e.~the planes on which ordering occurs. When SRO is present, this can be become complicated by the varying degrees of order within the chemical and structural heterogeneities within the dataset. To standardize this analysis, the Moran's I statistic for spatial autocorrelation, $M^I$, can be used to separately determine the type of ordering present in the Mg/Nb-site chemistry, the oxygen-oxygen distortions, and the oxygen-oxygen tilts. $M^I$ is calculated according to:

\begin{equation}
    M^I = \dfrac{N}{W} \dfrac{\sum\limits_{i=1}^{N}\sum\limits_{j=1}^{N}w_{ij}(A_i-\bar{A})(A_j-\bar{A})}{\sum\limits_{i=1}^{N}(A_i-\bar{A})^2}
\label{eq:morani}
\end{equation}

\noindent where

\begin{equation}
    W=\sum\limits_{i=1}^{N}\sum\limits_{j=1}^{N}w_{ij}
\label{eq:weights}
\end{equation}

\noindent and $w_{ij}$ are weights determined by the number of selected near neighbors, $A$ is the spatially-distributed feature of interest, such as normalized intensity, oxygen octahedral distortion, or oxygen octahedral tilt, and $N$ is the number of atom columns over which the feature is measured. This metric essentially quantifies the relative deviation of a value from its overall mean, evaluating to -1 when the measurements are identical in magnitude but opposite in sign in a particular $i,j$ neighborhood (negative numerator) and +1 when they are identical in magnitude and sign (positive numerator). When the measurement values are randomly distributed, $M^I$ is zero \citep{moran_1950, Cliff_Ord_Cliff_1981}. The type and degree of ordering present is then determined by comparing the $M^I$ statistic across increasing near neighbor shell index. This is conceptually similar to the  Warren-Cowley parameter where the probability of finding a specific atom type as a function of near neighbor shell indicates the type of chemical order \citep{cowley1950cuau}. Further, an associated Monte-Carlo test for statistical significance is performed in which the values across atom columns are scrambled and the Moran's I statistic re-evaluated to give a null distribution. 

\begin{figure*}
    \centering
    \includegraphics[width=3in]{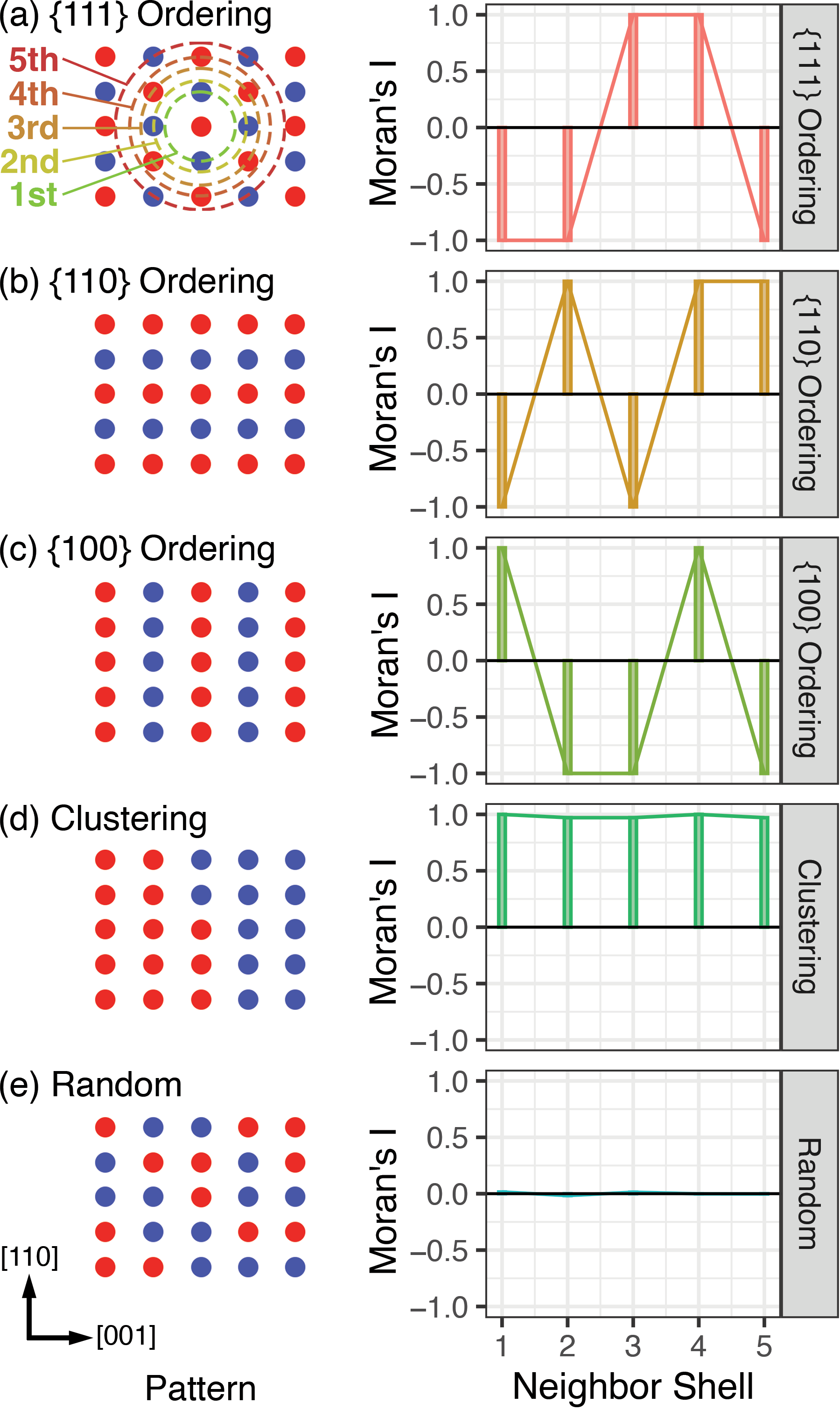}
    \caption{Types of ordering and corresponding Moran's I statistics across nearest like-neighbor shells: (a) $\{111\}$ order, (b) $\{110\}$ order, (c) $\{100\}$, (d) fully clustered, and (e) fully random example structures}
    \label{fig:globalmoran}
\end{figure*}

Using these tools, four test cases for $\{111\}$, $\{110\}$, and $\{100\}$ perfect ordering as well as complete clustering (segregation) and no order (random) are shown in Figure~\ref{fig:globalmoran}a-c. For the first five near neighbor shells,  $M^I$ shows distinct patterns across for each type. For example, perfect $\{111\}$ chemical order shows that near neighbors in the first, second, and fifth shells are completely dissimilar with a value of -1, while near neighbor atom columns in the third and fourth shells are identical, with a value of +1. Perfect $\{110\}$ ordering reveals a different pattern, with $M^I$ equal to -1 for the first and third shells and +1 for the other near neighbors. In the case of perfect $\{100\}$ ordering, $M^I$ is +1 for the first and fourth shells and -1 for the others. Finally, ordering can be separated from complete segregation (all $M^I$ = +1) or a random distribution (all $M^I$ = 0), as in Figure~\ref{fig:globalmoran}d,e.

\begin{figure*}
    \centering
    \includegraphics[width=3in]{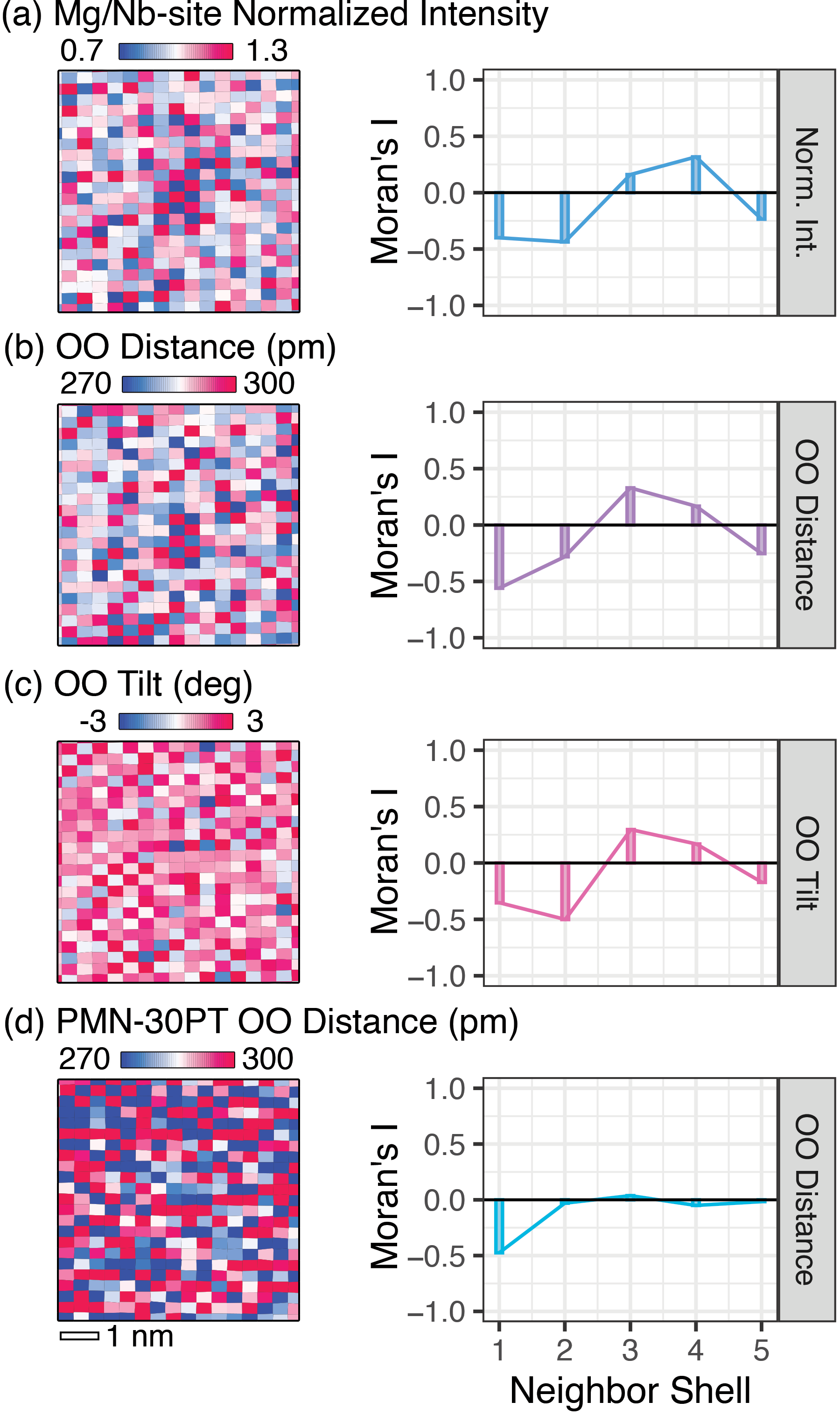}
    \caption{(a) Mg/Nb-site normalized intensity, (b) oxygen-oxygen distance, and (c) oxygen-oxygen tilt maps and their corresponding Moran's I statistics across near neighbor shells, showing an intermediate degree of $\{111\}$ order and random arrangement. (d) PMN-30PT oxygen-oxygen distance and the corresponding Moran's I statistics, showing mixed $\{111\}$, $\{110\}$, and random order. Each box in the maps represents one unit cell.}
    \label{fig:pmnglobal}
\end{figure*}

Applying $M^I$ analysis to the atomic-resolution STEM data of PMN allows for determination of the type of short-range order in the dataset. Based on the chemically sensitive normalized ADF intensity map of Mg/Nb-site atom columns, shown in Figure~\ref{fig:pmnglobal}a, the evaluated $M^I$ pattern of the B-site sub-lattice coincides with $\{111\}$ chemical ordering. Comparison the $M^I$ values to either perfect $\{111\}$ ordering or a random distribution (Figure~\ref{fig:globalmoran}a,e), however, the values of $M^I$ for PMN for each nearest like-neighbor shell are intermediate, reflecting the expected presence of both order and disorder within the dataset. This is apparent from visual inspection and also consistent with previous observations in PMN \citep{Hilton1990}. Similarly, oxygen-oxygen distances and tilts also exhibit a global tendency for $\{111\}$ ordering as shown in Figure~\ref{fig:pmnglobal}b,c. 

To demonstrate applicability to other types of ordering, we explore rhombohedral- and monoclinic-like oxygen octahedral distortions in Pb(Mg$_{1/3}$Nb$_{2/3}$)O$_{3}$-30 PbTiO$_{3}$ (PMN-30PT) \citep{singh2003, Singh_Pandey_Zaharko_2006}. The Moran's I autocorrelations yield an expected average of $\{111\}$ and $\{110\}$ ordering at the first, second, and third nearest like-neighbor shells (Figure~\ref{fig:pmnglobal}d). At the fourth and fifth neighbor shells, the autocorrelation values deviate from the average of $\{111\}$ and $\{110\}$ ordering and are closer to the random case. This tendency is due to the mixed and local nature of both rhombohedral- and monoclinic-like order, resulting in lower correlation at longer length scales \citep{kim2012}. Importantly, using the Moran's I autocorrelations at defined nearest like-neighbor shells, the type and relative degree of ordering can be measured objectively without manual input, allowing rapid multi- and cross-dataset measurements to be performed.

\subsection{Quantifying Local Order}

\begin{figure}
    \centering
    \includegraphics[width=\textwidth]{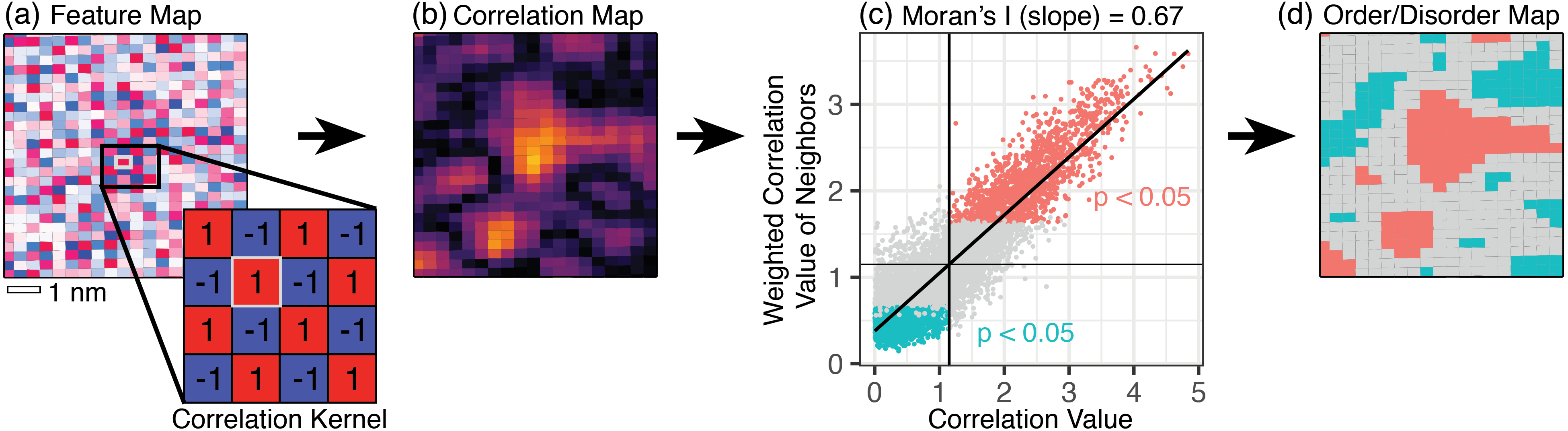}
    \caption{(a) Generic feature map showing checkerboard-alternating high and low values and a simple correlation kernel for template matching. (b) Output correlation map from (a) measuring the relative match to the correlation kernel. (c) Calculation of the global Moran's I statistic using comparison against weighted neighboring values using the correlation map, with low and high autocorrelation values indicating disorder and order, respectively. Statistically significant points determined with Monte-Carlo randomization are highlighted in red and blue. (d) Order/disorder map corresponding to unit cells categorized in (c). Each box in the maps represents one unit cell.}
    \label{fig:filtermoranlag}
\end{figure}

To quantify local order, a 4$\times$4  checkerboard kernel, shown in Figure~\ref{fig:filtermoranlag}a, is chosen so that the correlation for either completely uniform features, perfect $\{110\}$ ordering, or perfect $\{100\}$ ordering is zero. The $\{111\}$-type ordering determined by systematically evaluating the Moran's I values of near neighbor shells justifies the use of this checkerboard correlation kernel to locally average and highlight ordered clusters in the images \citep{kumar_2021pmn}. Furthermore, this kernel is applied to Mg/Nb-site normalized intensity, oxygen-oxygen distance, and oxygen-oxygen tilt maps with the resulting correlation maps emphasizing the $\{111\}$ ordered clusters (high correlation) and de-emphasizing disorder (low correlation) in chemistry and structure (Figure~\ref{fig:filtermoranlag}b).

Local cluster analysis is performed using a local indicator of spatial association, the Anselin Local Moran's I statistic, $M^I_i$:
\begin{equation}
    M^I_i = \dfrac{\sum\limits_{j=1, j\neq i}^{N}w_{ij}(A_i-\bar{A})(A_j-\bar{A})}{\sum\limits_{i=1}^{N}\dfrac{(A_i-\bar{A})^2}{n-1}}
\label{eq:localmorani}
\end{equation}
\noindent where $w_{ij}$ are weights determined by the number of selected near neighbor atom columns and $A$ is the spatially-distributed feature of interest, i.e.~intensity or distance.
This variation of the Moran's I statistic excludes the summation over all atom columns $i$, offering a local measure of high and low autocorrelation \citep{anselin}. Similar to the general Moran's I index, a Monte-Carlo randomization of feature values is automatically performed to obtain a distribution of $M^I_i$ expected under the null hypothesis. By comparing this null distribution with the measured value, a $p$-value is derived and adjusted for multiple hypothesis testing intrinsic to local cluster analysis \citep{Benjamini1995, Ord1995, CaldasdeCastro2006}. Here, the first and second nearest like-neighbors of the Mg/Nb sub-lattice atom columns define the ``neighborhood'' in which the local spatial autocorrelation is computed. 

The weighted average of neighboring correlated intensities is compared with the correlated intensity of each Mg/Nb-site atom column, resulting in the calculation and plot of the global Moran's I statistic (Figure~\ref{fig:filtermoranlag}c). High and low autocorrelation in the correlation maps are revealed by the points above the weighted and unweighted mean values, with a significance threshold of 5\%. These categories are then further defined as  highly  ordered (with $p<$0.05), highly  disordered (with $p<$0.05), and intermediate order regions (Figure~\ref{fig:filtermoranlag}d).

\begin{figure*}
    \centering
    \includegraphics[width=\textwidth]{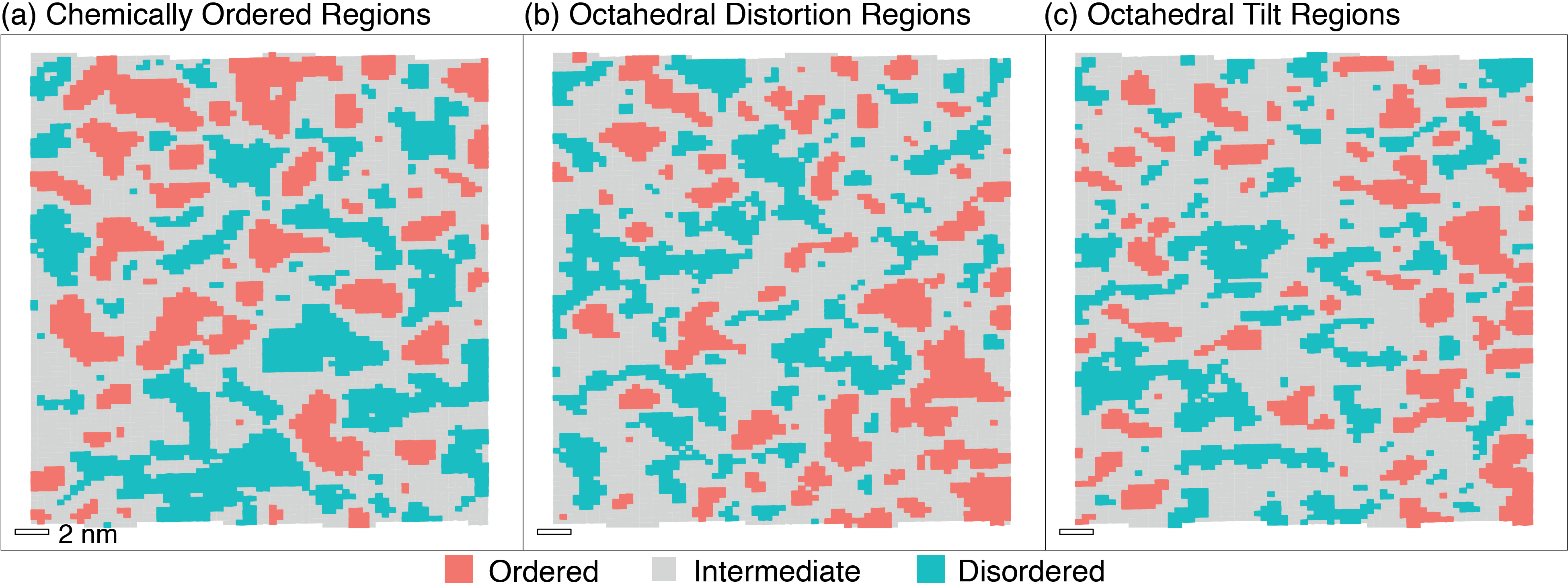}
    \caption{Local cluster analysis identifying order and disorder using the Anselin Moran's I. (a) Chemical order and disorder regions, (b)  octahedral distortion order and disorder regions, and (c) octahedral tilt order and disorder regions. Each box represents one unit cell. }
    \label{fig:localmoran}
\end{figure*}

With the categories defined, clusters of chemical and structural order are then mapped in Figure~\ref{fig:localmoran} and summarized in Table~\ref{table:ordersummary}.  Similar area fractions, cluster areas, correlation lengths, and separations are seen for all three features, with standard deviations marked where applicable. For example, the chemical ordering across the region, Figure~\ref{fig:localmoran}a, contains a relatively similar proportion of highly ordered and disordered regions separated by mixed regions of intermediate order/disorder. Additional spatial measurements show that the highly chemically ordered regions are widely scattered in size with an average area of 7.1 nm$^2$ and a correlation length of 1.5 nm, consistent with previously reported COR sizes of 2-6 nm measured by dark field TEM \citep{Krause_Cowley_Wheatley_1979, Hilton1990}. Furthermore, oxygen octahedral distortion and tilt order (Figure~\ref{fig:localmoran}b,c), representative of local rhombohedral structure, have correlation lengths and volume fractions in the range of 1-2 nm and 10-20\%, respectively, in agreement with results from neutron diffraction \citep{Xu_Shirane_Copley_Gehring_2004, Jeong_Darling_Lee_Proffen_Heffner_Park_Hong_Dmowski_Egami_2005}. 

\begin{table*}
\caption[Order Summary]{Summary of Local Order Across Chemistry, OO Distortion, and OO Tilt}
\begin{tabular}{p{0.3\textwidth} p{0.2\textwidth} p{0.2\textwidth} p{0.2\textwidth}}
\hline
Measure & Chemical Order & OO Distortion Order & OO Tilt Order\\
\hline

\% Highly Ordered  & 21 & 18  &  16 \\
\% Highly Disordered  & 25 & 19   & 18\\
\% Intermediate  & 54  & 63  & 66\\
Ordered Cluster Area (nm$^2$) & 7.1 ($\sigma=$ 4.9)  & 6.2 ($\sigma=$ 4.1)  & 4.9 ($\sigma=$ 2.9)\\
Correlation Length (nm) & 1.5 ($\sigma=$ 0.48) & 1.4 ($\sigma=$ 0.39) & 1.2 ($\sigma=$ 0.33)\\
Center Separation (nm) & 3.6 ($\sigma=$ 0.93) & 3.2 ($\sigma=$ 0.94) & 2.9 ($\sigma=$ 0.88)\\

\hline
\end{tabular}
\label{table:ordersummary}
\end{table*}

\subsection{Spatially-Correlated Chemistry and Structure}

Determining the spatial relationship between structural and chemical order can provide insight into further controlling overall material structure. Based on the statistically significant clusters of order and disorder mapped in Figure~\ref{fig:localmoran}, similarities in region properties exist across each feature type, concurring with with the linear correlation between global chemistry and structure (Figure~\ref{fig:intro}). Integration of local spatial context, however, offers an extension of previous analyses and sheds additional insight into regions of strong chemistry-structure correlation. 

Using a similar approach to spatial autocorrelation for identifying clusters of local order, the relationship between two features, such as chemistry and oxygen octahedral distortion, can also be evaluated. The codispersion coefficient measures this relationship and is defined as: 
\begin{equation}
    \rho_{AB}(h) = \dfrac{\hat{\gamma}_{AB}(h)}{\sqrt{\hat{\gamma}_{A}(h)\hat{\gamma}_{B}(h)}}
\label{eq:codispersion}
\end{equation}
\noindent where
\begin{equation}
    \hat{\gamma}_{AB}(h) = \dfrac{1}{2|N(h)|} \sum\limits_{N(h)}(A(x_i)-A(x_j))(B(x_i)-B(x_j))
\label{eq:crossvariogram}
\end{equation}
and $N(h)$ defines the cardinality, or neighborhood of paired observations of features $A$ and $B$ (such as chemical order and oxygen octahedral distortion order) in the distance group $h$. This statistic is based on both the semi-variogram, which describes the variance of a feature compared with itself at varying spatial distances \citep{Matheron_1963vario}, as well as the cross-variogram (Equation~\ref{eq:crossvariogram}), which describes the covariance of two different features as a function of distance. 
The codispersion coefficient effectively tracks co-movement of paired features in space by identifying how their covariance changes as a function of distance, allowing for both degree and extent of correlation to be determined \citep{vallejos2020, VerHoef_Barry_covario}.  For reference, features that are fully correlated spatially, yield a maximum value of one. 

\begin{figure}
    \centering
    \includegraphics[width=3in]{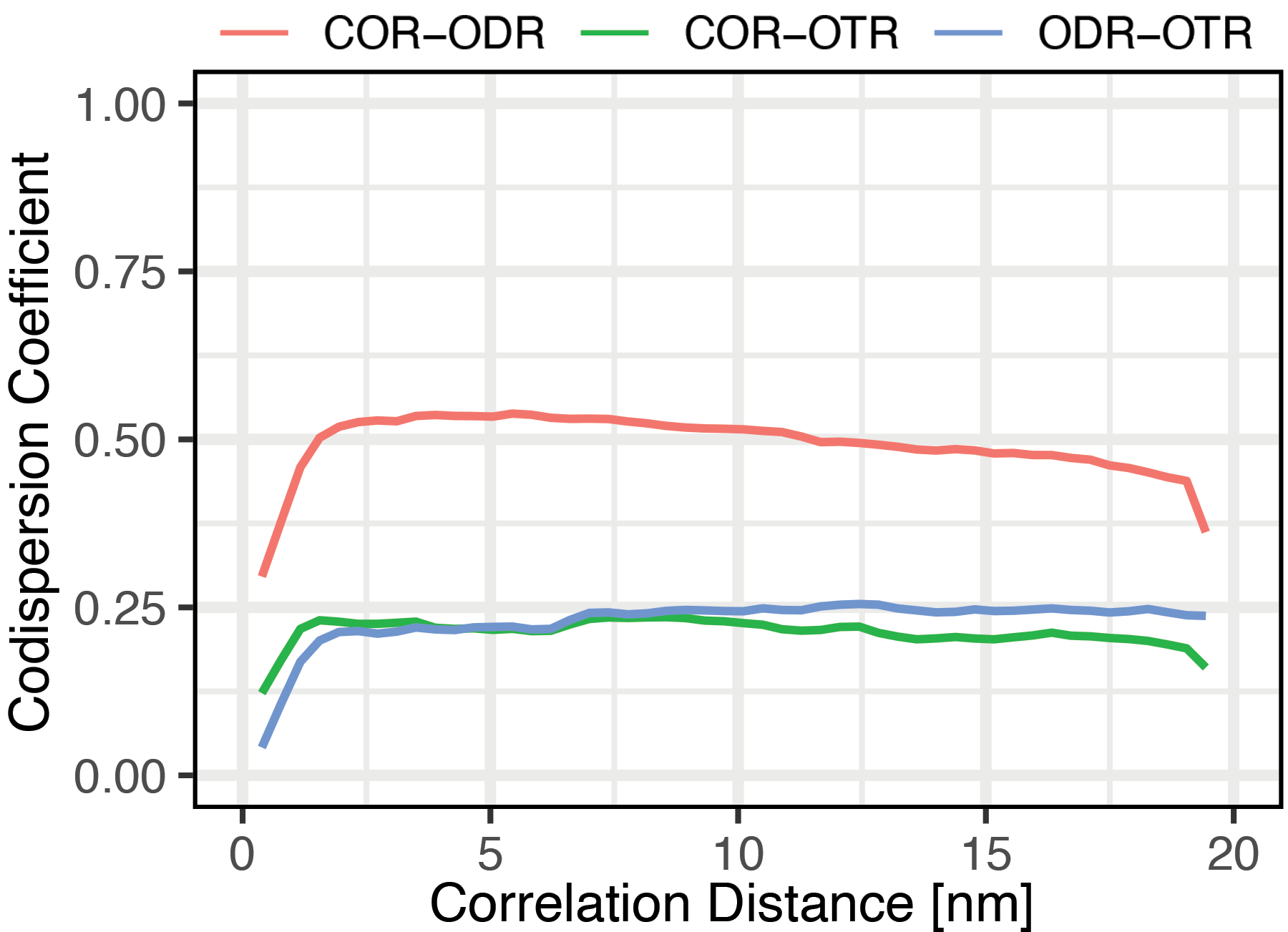}
    \caption{Codispersion coefficient as a function of spatial neighborhood (correlation distance) for chemistry, distortion, and tilt ordering. }
    \label{fig:codispersion}
\end{figure}

The codispersion relationship among chemical (COR), oxygen octahedral distortion (ODR), and oxygen octahedral tilt (OTR) ordering are shown for PMN in Figure~\ref{fig:codispersion}.Compared to the spatially-blind Pearson correlation coefficient, a continuum of correlation can be seen for various distance neighborhoods. A sharp increase in the codispersion coefficient, from the first near neighbor outward, indicates correlations with similar length scales as the individual autocorrelations. Additionally, a maximum in correlation distance between chemical order and oxygen octahedral distortion order occurs between 2-5 nm, representing the approximate length scale of strongly co-ordered clusters of combined chemical and structural distortion ordering. The codispersion coefficient remains relatively high ($\approx$0.5) at increased correlation distances,  resulting from similar length-scales for order and disorder of CORs and ODRs. At larger scales, however, codispersion coefficient decreases steadily as the overall spatial coincidence falls off.

The maximum codispersion coefficients are lower for chemical order-oxygen octahedral tilt order and oxygen octahedral distortion-tilt order, demonstrating weaker spatial correlation of these features. This is consistent with the corresponding global Pearson correlation coefficients. As the codispersion coefficient is essentially a normalized spatial covariance, the difference in magnitude among chemical-octahedral tilt and octahedral distortion-octahedral tilt heterogeneities indicates lower correlation, in other words the spatial features are more independent. Further, a nearly constant codispersion coefficient with increasing distance is present, reflecting a similar decreased correlation of order-disorder distributions even across a range of length scales compared to that of chemical order-oxygen octahedral distortion order. 

\begin{figure*}
    \centering
    \includegraphics[width=\textwidth]{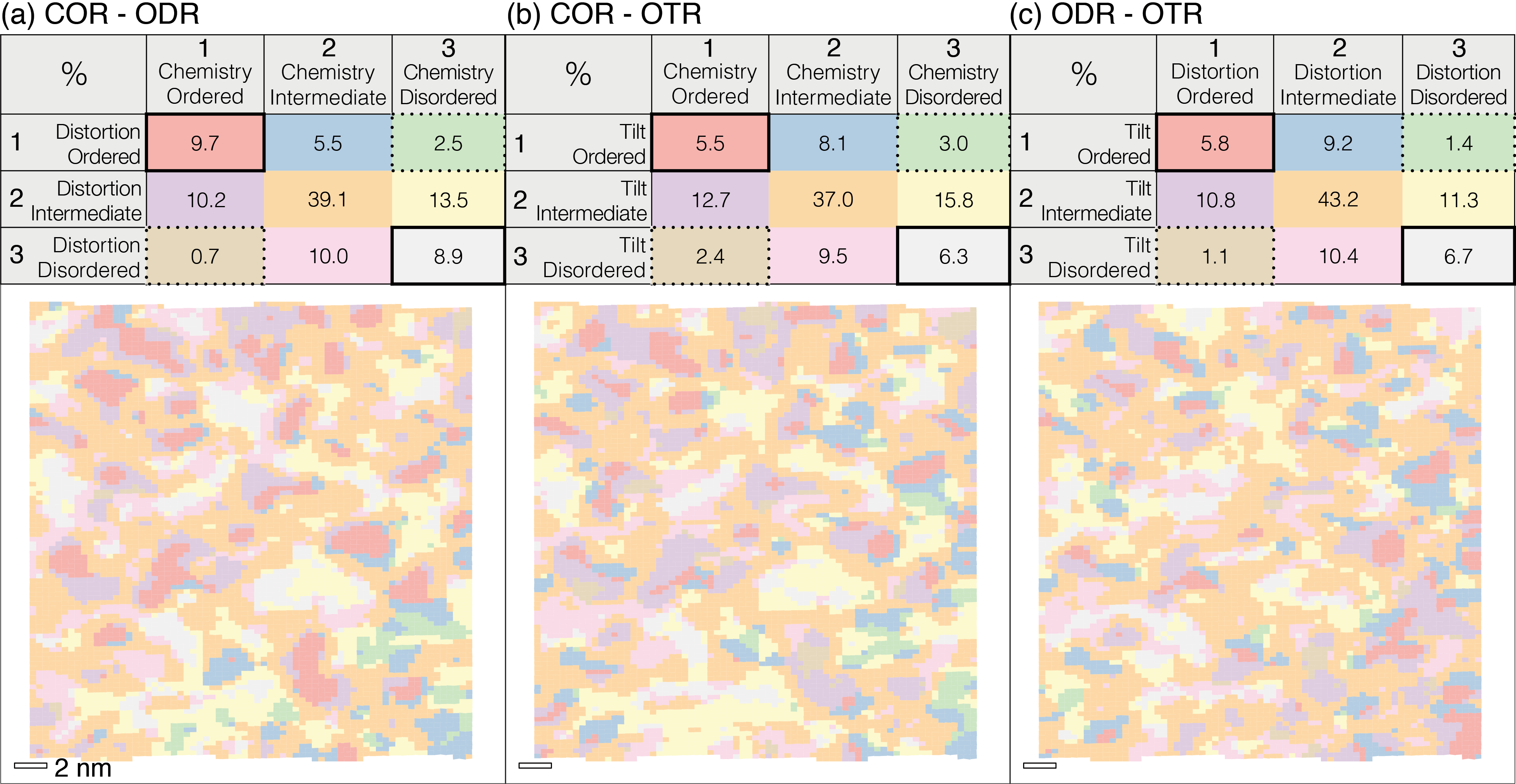}
    \caption{(a) Chemical-Distortion (COR-ODR), (b) Chemical-Tilt (COR-OTR), and (c) Distortion-Tilt (ODR-OTR) ordering spatial maps showing areas of correlation and anti-correlation based on overlap and statistical significance. Percentages of the total imaged area are given in the accompanying table, with  correlation bolded and anti-correlation dotted. }
    \label{fig:spatial_map}
\end{figure*}

Combined with codispersion coefficient analysis, order maps determined from the local Moran's I analysis can be used to visualize the extent of correlation in space. High order, intermediate order, and high disorder regions from Figure~\ref{fig:localmoran} are compared across CORs, ODRs, and OTRs to determine overlapping regions of spatial correlation. A visual interpretation of this correlated order distribution is shown in Figure~\ref{fig:spatial_map}. High order, high disorder, and intermediate order regions for any two features are simultaneously shown and their area fractions summarized in the tables above each map. For chemical order and oxygen octahedral distortion order in Figure~\ref{fig:spatial_map}a, the percentage of 1:1 positively-correlated regions across the entire dataset is approximately 57.7\% (i.e~sum along the table diagonal), similar in magnitude to the maximum codispersion coefficient calculated for those two features. This co-movement of chemical order and oxygen octahedral distortion order is visually confirmed at the local and global length scales by spatially-coincident high order (red), intermediate order (orange), and high disorder (light gray). Furthermore, the anti-correlated regions, marked by the dotted cells in the table, are minimal, indicating high spatial correlation. 

Moving away from strongly-correlated regions on the main diagonal (top left to bottom right), weakly-correlated regions indicated by the regions right above and below the main diagonal [table cells (1, 2), (2, 1), (2, 3), and (3, 2)] are present, in which high order is coupled with intermediate order. Moreover, anti-correlated regions are scarce, as represented by the percentages in the top right and bottom left cells. The presence of an extended gradient from correlation to anti-correlation confirms that the chemistry and distortion ordering phenomena are strongly but not fully correlated. 

Similar comparisons are made for chemistry-oxygen octahedral tilt order (Figure~\ref{fig:spatial_map}b) and oxygen octahedral distortion-tilt order (Figure~\ref{fig:spatial_map}c). While the area fraction of the intermediate-intermediate order in these two comparisons is similar to that of the chemistry-oxygen octahedral distortion relationship, differences are present in the directly- and anti-correlated features. As the percentage of coupled high order and high disorder regions (boxed and dotted cells, respectively) for chemical-tilt and distortion-tilt comparisons are approximately half that of chemistry-distortion, weaker spatial correlation between these features is seen, consistent with both codispersion analysis and previous Pearson correlation analysis. In addition, a greater amount of anti-correlated regions is present, supporting the notion of more independent phenomena governing these types of heterogeneities, particularly chemistry and oxygen octahedral tilt \citep{Rosenfeld_Egami_1995}. 


A gradient of correlation to anti-correlation is also present, with the percentage of mixed correlated regions off the diagonal of the table higher than with chemistry-distortion order. Significant coupling between chemistry and oxygen-octahedral distortion is thus evident by quantitative comparison of the spatial distribution of correlated order and disorder in combination with codispersion analysis. As both chemistry and oxygen octahedral distortion order have roles in disrupting long-range polarization \citep{kumar_2021pmn}, their spatial correlation indicates possibilities for control of polar domain structure and relaxor ferroelectric response.

\section{Conclusion}

We have shown that GIS-based spatial statistics methods can be used to determine correlations between chemistry and structural SRO in STEM imaging datasets. With PMN as a prototypical example, we establish, based on null-hypothesis testing, the regions defined by high order, intermediate order, and high disorder, with the resulting determination supported by previous studies. In addition, chemical order and oxygen octahedral distortion order are shown to be strongly-correlated, with spatial coincidence that extends across both ordered and disordered clusters. In contrast, only weak correlations between chemical-oxygen octahedral tilt order and oxygen octahedral distortion-tilt order are observed. Based on these results, we find that spatial correlation and autocorrelation can readily identify, categorize (ordered vs. disordered), and measure distance-based relationships between various types of SRO. Overall, by adopting spatial statistics tools commonly used for spatio-temporal studies in other fields, correlative relationships are measured and visualized to provide a wealth of information regarding short-range chemistry-structure interactions in PMN, with applicability to a variety of similarly complex material systems.

\section{Acknowledgements}

The authors thank Professor Shujun Zhang for providing the bulk PMN and PMN-PT samples. 
This work was supported by the Army Research Laboratory via the Collaborative for Hierarchical Agile and Responsive Materials (CHARM) under cooperative agreement W911NF-19-2-0119. This work made use of the MIT.nano Characterization Facilities. 

\normalsize

\bibliography{reference}
\end{document}